\documentclass[11pt,a4paper]{article}
\usepackage{jheppub} 
             
\title{Hydrodynamics in 1+1 dimensions with gravitational anomalies}

\author{Manuel Valle}

\affiliation{Departamento de F\'\i sica Te\'orica, 
Universidad del Pa\'is Vasco UPV/EHU, \\
Apartado 644,  48080 Bilbao, Spain}

\emailAdd{manuel.valle@ehu.es}

\abstract{
The constraints imposed  on hydrodynamics by the structure 
of gauge and gravitational anomalies are studied in two dimensions. 
By explicit integration of the consistent gravitational anomaly,  
we derive the equilibrium partition function at second derivative order.  
This partition function is then used to compute 
the parity-violating part of the covariant energy-momentum tensor 
and the transport coefficients. 
}
\keywords{}

\begin{document}
\maketitle
\flushbottom

\section{Introduction}
The recent studies of the role played by  gauge anomalies in relativistic hydrodynamics have 
revealed unexpected modifications in the constitutive relations of the anomalous currents 
\cite{Son:2009tf,Neiman:2010zi}.
These give rise to parity violating macroscopic phenomena such 
as the chiral magnetic and chiral vortical  effects \cite{Fukushima:2008xe, Landsteiner:2011iq}. 
It has also been pointed out in \cite{Landsteiner:2011cp} that the temperature corrections to 
the constitutive relations is related to the mixed gravitational anomaly.
The guiding principle which usually has served to derive the form of 
constitutive relations is the principle of entropy increase. 
However, as some of the new parity odd transport coefficients 
have even signature under time reversal, they produce non-dissipative effects 
and do not  contribute to entropy increase.  
Arguments based on the requirement of adiabaticity have just provided
a better understanding 
of anomaly induced effects \cite{Loganayagam:2011mu, Loganayagam:2012pz}, 
and they have helped as well to make progress in 
the  classification of the corrections in the derivative expansion to 
second order fluids dynamics \cite{Kharzeev:2011ds}. 

Very recently, it has been argued that it is possible to find 
all the corrections to the constitutive relations without dissipative coefficients 
from the knowledge of the equilibrium 
thermodynamics \cite{Jensen:2012jh,Banerjee:2012iz}.
This has been checked in various examples  \cite{Jensen:2012jh,Banerjee:2012iz,Jain:2012rh,Bhattacharyya:2012xi} 
where the form of the partition function 
is determined by general arguments of gauge and diffeomorphism invariance.  
 
In this note we consider the inclusion of 
the effects of the gravitational anomaly on the hydrodynamics in 1+1 dimensions, 
about which little is know. 
Previous studies of the effects of pure gravitational anomalies have focused on 
the connection of these effects with viscoelastic and thermal transport phenomena 
in topological insulators \cite{Ryu:2010fk,Stone:2012ud}. 
Here, our interest  is rather in the implications of gravitational anomalies  
in a relativistic setting, 
where a theory of chiral fermions is put in an external gravitational field. 
Specifically, we will construct an anomalous partition function in a curved manifold, and by following
the methods of \cite{Jensen:2012jh,Banerjee:2012iz}  derive the anomaly induced  constitutive relations. 
Our main finding is a novel  term  of second order in the derivatives of the velocity fluid 
which, in the Landau frame, shows up in the constitutive relation for current density. 
  
The  approach that we have followed for the computation of the partition function 
is based on the integration of the formula for the consistent gravitational anomaly 
in the special case of a time-independent background. 
As shown in section \ref{segunda}, in such a case the integration yields a local 
functional depending on the gravitational sources, 
but since the consistent anomaly is not generally covariant, 
the quantities obtained by functional differentiation with respect to the external sources 
do not transform like a tensor. 
As  it is well known,  the addition of a specific term known as 
Bardeen's polynomial \cite{Bardeen:1984pm} produces an 
energy-momentum tensor that is generally covariant, although no longer derivable from a partition function. 

With the partition function at hand, we apply the
systematic method presented in \cite{Jensen:2012jh,Banerjee:2012iz} for constraining 
the constitutive relations of hydrodynamics. 
It  is worth mentioning that the necessity  to shift from consistent to covariant currents 
is not peculiar to the gravitational case, as it already shows up in  \cite{Jensen:2012jh,Banerjee:2012iz,Jain:2012rh}, 
where only the role  of gauge anomalies is 
considered\footnote{For a careful discussion of the distinction between the role played of consistent and 
covariant currents, see also \cite{Banerjee:2012cr}.}. 
Finally, in section \ref{tercera}  the covariant 
energy-momentum tensor is used  to  
derive the modifications of the constitutive relations which follow from the gravitational anomaly.

\section{The gravitational anomaly and its contribution to the partition function}
\label{segunda}
In two dimensions the most general static  metric and gauge field  which are preserved by the Killing vector $\partial_t$ may 
be written as 
\begin{equation}\label{backg}
\begin{split}
ds^2 &= -e^{2 \sigma(x)} \left(dt + a_1(x) dx \right)^2 + g_{1 1} dx^2 ,   \\  
\mathcal{A} &= \mathcal{A}_0(x) dt + \mathcal{A}_1(x) dx . 
\end{split}
\end{equation}
We are interested in discussing the role of the gravitational anomaly 
in the hydrodynamics of a chiral charged fluid. 
The anomalous conservation law for the consistent energy-momentum 
tensor is  \cite{AlvarezGaume:1983ig, Bardeen:1984pm}
\begin{equation}
\nabla_\mu T^{\mu \nu} =  
D\, \epsilon^{\gamma \delta} \partial_\alpha \partial_\gamma  \Gamma_{\delta \rho}^{\;\;\;\alpha} \, g^{\rho \nu} ,
\end{equation}
where $\epsilon^{01} = \frac{1}{\sqrt{-g}} = \left(e^{\sigma}\sqrt{g_{11}}\right)^{-1}$, 
and the coefficient  $D$ is $\pm 1/(96 \pi)$ for a Weyl fermion. 
The fact that this tensor is constructed from the variation of a functional action $W$  
constrains the form of the anomalous divergence through a 
consistency condition which follows from 
the commutation relations between infinitesimal coordinate transformations  \cite{Bardeen:1984pm}.
Such a transformation is specified by  infinitesimal parameters $\xi^\mu(x)$, 
and  may be  decomposed into two parts when acting on non-tensorial quantities, 
$\delta_\xi =   \mathcal{L}_\xi + \delta_\Lambda$.  
For instance, by using  matrix notation for the vielbein and the connection,  $(E)_\mu^{\; \; a}\equiv e_\mu^{\; \; a}$, 
$(\Gamma_\lambda)_\mu^{\;\; \nu}\equiv \Gamma_{\lambda \mu}^{\quad \nu}$, 
these transform as \cite{Bardeen:1984pm}
\begin{equation}
\label{delta}
\begin{split}
\delta_\xi E &=  \mathcal{L}_\xi  E  - \Lambda E, \\
\delta_\xi \Gamma_\lambda &= \mathcal{L}_\xi  \Gamma_\lambda +\left[\Gamma_\lambda, \Lambda\right], 
\end{split}
\end{equation}
where $\mathcal{L}_\xi$ is the Lie derivative 
and  $\Lambda$ is the matrix  $(\Lambda)_\mu^{\; \; \nu} \equiv -\partial_\mu \xi^\nu$. 
This $\Lambda$ may be viewed as the matrix  specifying a gauge transformation on the connection. 
It is possible to express the effective action $W$  by introducing a set of fields $H$ which transform non-linearly 
under the gauge part $\delta_\Lambda$ for the connection in \eqref{delta}. 
These fields $H$ are given in terms of the vielbein by the matrix relation $e^H = E$. 
Therefore, the effective anomalous action is given by \cite{Bardeen:1984pm,Hwang:1985uj}
\begin{equation}
\label{wanom}
W_\mathrm{anom}[H, \Gamma] =  
D \int d^2 x \sqrt{-g}\int_0^1 ds\, \mathrm{Tr}\left(H\, \partial_\rho \Gamma_\lambda(s) \epsilon^{\rho \lambda}\right) ,
\end{equation}
where 
\begin{equation}
\Gamma_\lambda(s) = e^{-s H} \Gamma_\lambda e^{s H} + e^{-s H} \partial_\lambda e^{s H}  . 
\end{equation} 

For the background metric \eqref{backg}  the integration over $s$ may be explicitly performed.   
The result is expressed, remarkably, as  a local functional of the  
metric fields $\sigma(x), a_1(x)$, $g_{11}(x)$ 
and their derivatives up to second order  
\begin{equation}
W_\mathrm{anom}^{(2)} = \int dt \,  \int dx \, P_\mathrm{anom} ,  
\end{equation}
where the integrand turns out to be 
\begin{equation}\label{anom}
\begin{split}
\frac{1}{D} \frac{1}{\sqrt{-g}} P_\mathrm{anom}&=
\frac{e^{2 \sigma} a_1 \sigma''}{2 g_{11} \left(e^\sigma - \sqrt{g_{11}}\right) } \left(\ln g_{11} - 2 \sigma \right)\\
&\quad+\frac{e^{2 \sigma} a_1 \left(g_{11}' \sigma' -2 g_{11} \sigma'^2 \right)}{4 g_{11}^2 \left(e^\sigma - \sqrt{g_{11}}\right)^2} \\
&\qquad \times \bigl[
          e^\sigma(2-\ln g_{11} + 2 \sigma) +2 \sqrt{g_{11}}(-1+\ln g_{11} -2 \sigma)  \bigr]  \\ 
&\quad + \frac{e^{2 \sigma} a_1' \sigma'}{2 g_{11}^{3/2} \left(e^\sigma - \sqrt{g_{11}}\right)}
\bigl[2 e^\sigma + \sqrt{g_{11}} (-2+\ln g_{11} - 2 \sigma) \bigr] . 
\end{split}
\end{equation}
This locality  of the anomalous effective action has also been found 
in \cite{Loganayagam:2011mu, Loganayagam:2012pz,Banerjee:2012iz, Jain:2012rh},  
where consistent static currents were derived from  a local contribution to the partition function. 
From a diagrammatic point of view,  the origin of the locality in the static case can be traced to the fact 
that restricting the amplitudes to zero frequency 
turns the rational dependence on  the external momenta  into  a 
polynomial one. 
 
Since at thermal  equilibrium with inverse temperature $\beta_0$ the euclidean vacuum functional $W$ is 
related with the thermodynamic potential $\Omega$ through $W \equiv  -\beta_0 \Omega$, 
it follows that the function $P_\mathrm{anom}$ represents an anomalous contribution to the pressure. 
Following the notation in \cite{Banerjee:2012iz}, we write the anomalous contribution to the partition function as 
\begin{equation}
W_\mathrm{anom}^{(2)}=   \frac{1}{T_0} \int dx\,  P_\mathrm{anom} =  -\beta_0 \Omega[g_{\mu \nu}] . 
\end{equation} 
The components of the consistent energy-momentum tensor are easily obtained by functional differentiation
\begin{equation}
T^{\mu \nu} = \frac{2 T_0}{\sqrt{-g}} \frac{\delta W}{\delta g_{\mu \nu}(x)} = 
          -\frac{2}{\sqrt{-g}} \frac{\delta \Omega}{\delta g_{\mu \nu}(x)}  ,             
\end{equation}
yielding the following simple expressions 
\begin{align}
\label{bad1}
T_{00} &= -\frac{T_0 e^{2 \sigma}}{\sqrt{-g}} \frac{\delta W}{\delta \sigma}=
D \frac{e^{3 \sigma}}{g_{11}^{5/2}}\left(-a_1' g_{11}'  + g_{11} a_1'' \right) , \\ 
T_{0}^1&=  \frac{T_0}{\sqrt{-g}} \frac{\delta W}{\delta a_1}= D \frac{e^\sigma}{g_{11}^{5/2}}\left(g_{11}' \sigma' -2 g_{11}\sigma'^2 - g_{11} \sigma'' \right) , \\ 
\label{bad3}
T^{11} &=\frac{2 T_0}{\sqrt{-g}} \frac{\delta W}{\delta g_{11}}
= -D \frac{2 e^\sigma}{g_{11}^{5/2}} a_1' \sigma' . 
\end{align}

We have computed the components that, according to \cite{Banerjee:2012iz},  should be invariant 
under Kaluza-Klein gauge transformations. 
These transformations are redefinitions of time, $t \to t' = t + \phi(x)$, without change in the spatial coordinate,  
that preserve the form of the metric if $a_1$ transforms  as $\delta a_1 = -\phi'$. 
As discussed in \cite{Banerjee:2012iz}, the invariant combinations of the 
components of a tensorial quantity $V^{\mu \nu}$ correspond to 
 the purely spatial indices $V^{11}$, the lower temporal indices $V_{00}$, and $V_0^1$. 
 Clearly,  the above components $T_{00}$ and $T^{11}$  do not satisfy this property because of their specific dependence  on the 
 derivatives of $a_1$. 
 
This can be traced to the fact that, under a spatial diffeomorphism, 
the consistent components that  we have computed do not transform as tensorial quantities. 
Fortunately, there is a covariant modification $\widetilde{T}^{\mu \nu}$  which may be obtained through a shift $Y^{\mu \nu}$
 of the original combination \cite{Bardeen:1984pm},
 \begin{equation}
 \widetilde{T}^{\mu \nu} =  T^{\mu \nu} + Y^{\mu \nu}. 
 \end{equation}
This is called the covariant energy-momentum tensor and its divergence will also be covariant. 
It is given by \cite{Bardeen:1984pm, AlvarezGaume:1983ig}
\begin{equation}
\nabla_\mu \widetilde{T}^{\mu \nu} =  
-D\, \epsilon^{\nu \rho} \partial_\rho R ,
\end{equation}
where $R$ is the scalar curvature,
\begin{equation}
R = \frac{1}{g_{11}^2} \left( g_{11}' \sigma' - 2 g_{11} \sigma'^2 - 2 g_{11} \sigma'' \right). 
\end{equation} 
It is possible to show that,  in general,  $Y^{\mu \nu}$ must be a symmetric local quantity depending on 
second derivatives of the metric \cite{Bardeen:1984pm}.  
For our background metric we find 
\begin{align}
\frac{1}{D} Y^{00} &=\frac{e^{-\sigma}}{g_{11}^{5/2}}  \bigl[a_1' \left(g_{11}' + 2 e^{2 \sigma} a_1^2 \sigma' \right)  
   - g_{11} (a_{11}'' -4 a_1 \sigma'^2 + 2 a_1 \sigma'') \bigr] ,   \\ 
\frac{1}{D} Y^{01} &= \frac{e^{-\sigma}}{g_{11}^{5/2}}  \bigl[-2 e^{2\sigma} a_1 a_1' \sigma' - 2 g_{11} \sigma'^2 + g_{11} \sigma'' \bigr]  , \\
\frac{1}{D} Y^{11} &=  \frac{2 e^\sigma}{g_{11}^{5/2}} a_1' \sigma'  . 
\end{align}
With these results  at hand,  we finally  obtain the  anomalous covariant energy-momentum tensor. 
Their components are given by 
\begin{align}
\frac{1}{D}\widetilde{T}^{00}&= \frac{2 e^{-\sigma} a_1}{g_{11}^{5/2}} \left( g_{11}' \sigma' - 2 g_{11} \sigma'' \right), 
 &  \widetilde{T}_{00} &= 0, \\ 
\frac{1}{D} \widetilde{T}^{01} &= -\frac{e^{-\sigma}}{g_{11}^{5/2}}  \left( g_{11}' \sigma' - 2 g_{11} \sigma'' \right) , 
\label{t01}
& \frac{1}{D} \widetilde{T}_0^1 &= \frac{e^{\sigma}}{g_{11}^{5/2}}  \left(g_{11}' \sigma' - 2 g_{11} \sigma'' \right) , \\
{}& {} & \widetilde{T}^{11}&= 0 . 
\end{align}

The gauge invariance of  $\widetilde{T}^{\mu \nu}$ 
with respect to Kaluza-Klein gauge transformations, 
absent in \eqref{bad1} and \eqref{bad3},  is now manifest. 
It is also worthwhile checking the tensorial character of 
$ \widetilde{T}^{\mu \nu}$.  Under a diffeomorphism generated by $\xi^\mu = (0, \xi(x))$  
the Lie derivative of the metric with respect  to $\xi$  produces  the changes
\begin{align}
\delta \sigma(x) &= \xi(x) \sigma' (x), \\ 
\delta a_1(x) &= \xi(x) a_1'(x)  + \xi' (x)a_1(x), \\ 
\delta g_{11}(x) &= \xi(x) g_{11}' + 2 \xi'(x) g_{11} , 
\end{align}
which may be used to compute the induced change $\delta \widetilde{T}^{\mu \nu}(x)$. 
The calculation of these components using the above rules 
shows that $\delta \widetilde{T}^{\mu \nu}(x)$ precisely coincides
with the expression for $\mathcal{L}_{\xi}   \widetilde{T}^{\mu \nu}$. 
This provides a  non-trivial check of the whole computation because, in principle,   
 $\delta \widetilde{T}^{\mu \nu}(x)$ receives contributions proportional to $\xi''(x)$ which cancel out in the 
 final result. 
 
In order to complete the description of an anomalous charged fluid, 
we must also include the effects of the gauge anomaly,  
which already show up at zero derivative order. 
The consequences of the gauge anomaly on the hydrodynamics in 1+1 
dimensions have been studied in \cite{Dubovsky:2011sk, Jain:2012rh}.
The most general partition function
which encodes the effect of the gauge anomaly at zero order in 
the derivative expansion takes the form \cite{Jain:2012rh}
\begin{equation}\label{zeroorder}
W^{(0)} = W_\mathrm{anom}^{(0)}  + W_\mathrm{inv}^{(0)}  = 
-\frac{C}{T_0} \int \mathcal{A}_0\left(\mathcal{A}_1 - \mathcal{A}_0 a_1\right) dx 
- C_2 T_0 \int a_1 dx , 
\end{equation}  
where $C$ is the coefficient in the consistent anomaly, 
$\nabla_\mu J^\mu = C \epsilon^{\mu \nu} \partial_\mu \mathcal{A}_\nu$,   
and $C_2$ is an arbitrary coefficient. 
Hence $W_\mathrm{anom}^{(0)} +W_\mathrm{anom}^{(2)} $ encompasses the effects of the anomalies 
in the background \eqref{backg}. 
This action produces the  consistent current 
\begin{equation}
J^\mu = \frac{T_0}{\sqrt{-g}} \frac{\delta W}{\delta \mathcal{A}_\mu}, 
\end{equation}
that, when shifted by $C \epsilon^{\mu \lambda} \mathcal{A}_\lambda$, yields 
the gauge covariant current 
\begin{equation}
\widetilde{J}^\mu =  J^\mu + C \epsilon^{\mu \lambda} \mathcal{A}_\lambda , \qquad 
\nabla_\mu \widetilde{J}^\mu = C \epsilon^{\mu \nu}  \mathcal{F}_{\mu \nu} . 
\end{equation}
Its components are given by \cite{Jain:2012rh}
\begin{equation}\label{jtotal}
 \tilde{J}_0 = 0, \qquad  \tilde{J}^1 =  -2 C \frac{e^{-\sigma}}{\sqrt{g_{11}}} \mathcal{A}_0 .  
\end{equation}

One immediate consequence of the addition of $W^{(0)}$  to $W^{(2)}$ is that now  
the non-conservation law for the consistent energy-momentum ``tensor'' adopts the form 
\begin{equation}
\nabla_\mu T^{\mu \nu} =  
D\, \epsilon^{\gamma \delta} \partial_\alpha \partial_\gamma \Gamma_{\delta \rho}^{\;\;\;\alpha} \, g^{\rho \nu}  + 
\mathcal{F}^{\nu \alpha} \widetilde{J}_\alpha , 
\end{equation}
as follows from the behavior of the total functional $W = W^{(0)} + W^{(2)}$ under diffeomorphisms. 
Thus the conservation law for the covariant counterpart reads
\begin{equation}
\nabla_\mu \widetilde{T}^{\mu \nu} =  
-D\, \epsilon^{\nu \rho} \partial_\rho R +  
\mathcal{F}^{\nu \alpha} \widetilde{J}_\alpha . 
\end{equation}
By adding to \eqref{t01} the zero order contribution which follows from  \eqref{zeroorder},
the parity odd part of the stress tensor reads
\begin{equation}\label{t01total}
\widetilde{T}_0^1 = \frac{e^{-\sigma}}{\sqrt{g_{11}}} \left(C \mathcal{A}_0^2 -T_0^2 C_2 \right) + 
D \frac{e^{\sigma}}{g_{11}^{5/2}}  \left(g_{11}' \sigma' - 2 g_{11} \sigma'' \right) , 
\end{equation} 
while the remaining Kaluza-Klein gauge invariant components $\widetilde{T}_{00}$ and $\widetilde{T}^{11}$ vanish.

\section{Anomaly induced transport coefficients in the Landau frame}
\label{tercera}
We now seek the form of the constitutive relations that arise from  \eqref{jtotal}  and \eqref{t01total}.
At zero order in the derivative expansion, the time-independent equilibrium fluid fields in 
the background \eqref{backg} are given by 
\begin{equation}\label{equil}
\begin{split}
u_K^\mu &= e^{-\sigma}(1, 0), \\
T &= T_0 e^{-\sigma}, \\ 
\mu & = \mathcal{A}_0 e^{-\sigma} , 
\end{split}
\end{equation}
where $\mu$ denotes the chemical potential. Generally, 
as explained in \cite{Banerjee:2012iz} and \cite{Jensen:2012jh},  
these quantities receive derivative corrections 
which  can be expressed in terms of some specific combinations 
of covariant derivatives of the background data. Such combinations are non zero at equilibrium and can be determined from
the knowledge of the partition function. The derivative terms of the constitutive relations that arise in this way  are 
static susceptibilities rather than  transport coefficients related to genuine irreversible processes. 

In 1+1  dimensions the velocity field may also receive corrections of zero order in the derivative expansion \cite{Jain:2012rh}. 
This is due to the existence of a non-zero independent  vector field  
\begin{equation}\label{ueps}
\tilde{u}_K^\mu \equiv \epsilon^{\mu \nu} u_{K \nu} = 
\left(-\frac{a_1}{\sqrt{g_{11}}}, \frac{1}{\sqrt{g_{11}}}\right), \qquad \tilde{u}_{K 0} = 0, \qquad \tilde{u}_{K} ^1= \frac{1}{\sqrt{g_{11}}},
\end{equation}
which may be constructed by Hodge duality. 
On the other hand, the second order derivative contribution in \eqref{t01total} contains  
a quantity that may be written as 
the Laplacian on $\sigma$ with respect to the one-dimensional metric $g_{11}$ 
\begin{equation}
\nabla^2\sigma = \frac{1}{\sqrt{g_{11}}} \frac{d}{d x} \left(\sqrt{g_{11}} \frac{1}{g_{11}} \frac{d\sigma}{dx} \right) = 
 \frac{1}{2 g_{11}^2}  \left(-g_{11}' \sigma' + 2 g_{11} \sigma'' \right) . 
\end{equation}
If one considers an arbitrary normalized fluid velocity,  
this term is precisely  the second order scalar 
$(u^\mu \nabla^\rho - u^\rho \nabla^\mu)\nabla_{\mu} u_{\rho}$ for  $u^\mu = u_K^\mu$
\begin{equation}
 (u_K^\mu \nabla^\rho - u_K^\rho \nabla^\mu)\nabla_{\mu} u_{K \rho} = \nabla^2\sigma .
\end{equation}
Therefore, using \eqref{equil} and \eqref{ueps},  we can  express the anomalous contributions \eqref{jtotal}  
and \eqref{t01total} 
in terms of the fluid fields and their derivatives evaluated at equilibrium
\begin{equation}
\begin{split}
J_{\mathrm{anom}}^\mu &= -2 C \mu \, \tilde{u}^\mu, \\ 
T_{\mathrm{anom}}^{\mu \nu} &=  \bigl[C_2 T^2 - C \mu^2 +  
2D (u^\mu \nabla^\rho - u^\rho \nabla^\mu)\nabla_{\mu} u_{\rho} \bigr]\left(u^\mu \tilde{u}^\nu + u^\nu \tilde{u}^\mu\right) .
\end{split}
\end{equation}
These expressions are precisely  the parity odd corrections  to 
the constitutive relations. 
Note that they are
expressed in the frame defined by the Killing direction of the metric  \eqref{backg}
used to compute the anomalous thermodynamics. 
Together with  the even contribution from the zero derivative partition function,   
they imply the following forms for the current and the energy-momentum tensor
\begin{equation}\label{thermo}
\begin{split}
J^\mu &= n u^\mu + J_{\mathrm{anom}}^\mu, \\ 
T^{\mu \nu} &= \varepsilon u^\mu u^\nu + p P^{\mu \nu} + T_{\mathrm{anom}}^{\mu \nu}, 
\end{split}
\end{equation}
where $P^{\mu \nu} \equiv g^{\mu \nu} + u^\mu u^\nu$ and 
$p, \varepsilon, n$ are, respectively,  the pressure, energy and charge densities 
which follow from the zero order partition function.

In contrast, in 1+1 dimensions the tensor corrections to the perfect fluid part of the energy-momentum 
tensor necessarily  vanish  in 
the Landau frame\footnote{The three constraints $\Delta T^{\mu \nu} u_\nu = 0$, 
$g_{\mu \nu} \Delta T^{\mu \nu} = 0$,  define a vanishing  $\Delta T^{\mu \nu}$.}, 
so that all the modifications to the constitutive relations appear in the current $J^\mu$,  
and perhaps also in the pressure 
\begin{equation}\label{landau}
\begin{split}
J^\mu &= n u^\mu +  \chi\,   \tilde{u}^\mu , \\ 
T^{\mu \nu} &= \varepsilon u^\mu u^\nu + (p + \Delta p) P^{\mu \nu}  . 
\end{split}
\end{equation} 
The transformation to this frame may be written as
\begin{equation}
u^\mu =   u_K^\mu  \cosh \gamma +  \tilde{u}_K^\mu \sinh \gamma , 
\end{equation}
while the temperature and chemical potential will not receive corrections at  
linear order in the anomaly coefficients  $C, D$ and $C_2$. 
Plugging this equation into \eqref{landau} and comparing with \eqref{thermo} the parameter 
 $\gamma$ may be  determined. 
 One finds 
\begin{equation}
\gamma = \frac{1}{\varepsilon + p} \left[C_2 T^2 - C \mu^2 +  
2D (u^\mu \nabla^\rho - u^\rho \nabla^\mu)\nabla_{\mu} u_{\rho} \right] +
\text{second order corrections}. 
\end{equation}
This gives the anomaly induced coefficients of the constitutive relations in the Landau frame 
\begin{equation}
\begin{split}
 \chi &= -C_2 \frac{n T^2}{\varepsilon + p} + C \left(\frac{n \mu^2}{\varepsilon + p} -2 \mu\right) - 
 2 D\frac{n (u^\mu \nabla^\rho - u^\rho \nabla^\mu)\nabla_{\mu} u_{\rho} }{\varepsilon + p} , \\ 
  \Delta p &= 0 . 
\end{split}
\end{equation}

Note that  there are also parity even contributions at two-derivative order which we have not computed.
Without knowledge of the specific form of the second order partition function these corrections are undetermined.

\section{Conclusion}
\label{final}
In this note we have computed, in 1+1 dimensions,  the contribution of the gravitational anomaly to the partition function 
in the most general background with external sources consistent with thermodynamic equilibrium. 
The energy-momentum tensor at equilibrium has been evaluated as a variation of the 
partition function and then corrected by the appropriate term to make it generally covariant. 
By following the methods of \cite{Banerjee:2012iz} and \cite{Jensen:2012jh}, 
we have obtained the corresponding contribution to the constitutive relations 
without resort to the principle of entropy increase. 
The coefficient of the temperature correction does not seem to be determined 
by  the  anomaly coefficients, and it appears as an independent quantity.  

It would be interesting to perform a similar treatment in order to study   
the effects of the mixed 
gauge-gravitational anomalies in 3+1 dimensions up to third order in the derivative expansion.

\section*{Acknowledgements}
I would like to thank Juan L. Ma\~nes for 
many helpful discussions. 
This work was supported by funds from 
the Spanish Ministry of Science and Technology 
under grant FPA2009-10612, the Consolider-Ingenio Programme CPAN (CSD2007-00042),  
and  the Basque Government under grant IT559-10.

\bibliographystyle{JHEP}
\bibliography{bibgravanom}

\end{document}